\documentclass{article}
\usepackage{amssymb,amsmath}
\numberwithin{equation}{section}
\newtheorem{theorem}{Theorem}[section]
\newtheorem{remark}{Remark}
\newtheorem{lemma}{Lemma}
\begin{document}
\title{Existence of initial data satisfying the constraints for the 
spherically symmetric \\ Einstein-Vlasov-Maxwell system}
\author{P. Noundjeu$^{1}$; N. Noutchegueme$^{1}$; A. D. Rendall$^{2}$ \\ 
$^{1}$Department of Mathematics, Faculty of Sciences, University of 
Yaounde 1,\\ Box 812, Yaounde, Cameroun \\ {e-mail: noundjeu@uycdc.uninet.cm, 
nnoutch@uycdc.uninet.cm} \\ $^{2}$Max-Planck-Institut f\"ur 
Gravitationsphysik, Am M\"uhlenberg 1, \\14476 Golm, Germany \\ 
{e-mail: rendall@aei.mpg.de} }
\date{}
\maketitle
\begin{abstract}
Using ODE techniques we prove the existence of large classes of initial data
satisfying the constraints for the spherically symmetric 
Einstein-Vlasov-Maxwell system. These include data for which the ratio of
total charge to total mass is arbitrarily large.
\end{abstract}
\section{Introduction}

The global dynamical behavior of self-gravitating matter is a subject of
central importance in general relativity. A form of matter which has
particularly nice mathematical properties is collisionless matter, described
by the Vlasov equation. It has the advantage that it lacks the tendency 
observed in certain other models, such as perfect fluids, that solutions
of the equations of motion of the matter lose differentiability after a 
finite time. These singularities of the mathematical model form an obstacle
to further analysis and prevent the study of the global dynamical 
properties of the solutions. Collisionless matter is free from these
difficulties and there is a growing literature on global properties of
solutions of the Einstein-Vlasov system \cite{andreasson}, \cite{rendall1}.

In \cite{rein}, the authors prove the global existence of asymptotically 
flat solutions of the spherically symmetric Einstein-Vlasov system, with 
small initial data. That study concerns uncharged particles. We consider, 
under the same assumption of spherical symmetry, the case where the 
particles are charged. To describe the full physical situation, we must
then couple the previous system to the Maxwell equations that
determine the electromagnetic field created by the fast moving
charged particles. As will be seen below, that reduces, in the spherically 
symmetric  case, to its electric part. 

It is appropriate at this point to examine the motivation for considering
this particular problem. We are not aware that it has any direct 
astrophysical applications. There are, however,  two reasons why the problem 
is interesting. The first is that it extends the knowledge of the Cauchy 
problem for systems involving the Vlasov equation and it will be seen that 
it gives rise to new mathematical features compared to those cases studied 
up to now. The second is connected with the fact that it would be desirable 
to extend the work of \cite{rein} beyond spherical symmetry. In particular, 
it would be desirable from a physical point of view to include the 
phenomenon of rotation. Unfortunately, presently available techniques do 
not suffice to get away from spherical symmetry. In this situation it is 
possible to attempt to obtain further intuition by using the analogy between 
angular momentum and charge, summed up in John Wheeler's statement, 
\lq Charge is a poor man's angular momentum\rq. Thus we study spherical 
systems with charge in the hope that this will give us insight into 
non-spherical systems without charge. This strategy has recently been 
pursued in the case of a scalar field as matter model, with interesting 
results \cite{dafermos1}, \cite{dafermos2}.

Concerning the Cauchy problem in general relativity, it is well known that 
in addition to the Einstein evolution equations there are constraint 
equations which have to be solved. (See e.g. \cite{friedrich}.) It is only of 
interest to consider the problem of evolution, if the problem of constraints 
on the initial data can be solved. In our specific case, we are led to a 
difficulty in solving the constraints on the initial data, that has not 
previously been considered in the literature. Let us first recall the 
situation in \cite{rein} before seeing how it changes in the case of charged 
particles. In \cite{rein}, using the assumption of spherical symmetry, the 
authors look for two metric functions $\lambda$ and $\mu$, that depend only 
on the time coordinate $t$ and on the radial coordinate $r$, and for the 
distribution function $f$ of the uncharged particles that depends on $t$, $r$
and on the 3-velocity $v$ of the particles; the metric functions
$\lambda$, $\mu$ are subject to the Einstein equations with
sources generated by the distribution function $f$ of the
collisionless uncharged particles which is itself subject to the Vlasov
equation. They show that the Einstein equations to determine
the unknown metric functions $\lambda$ and $\mu$, turn out to be
two first order ODE in the radial variable $r$, coupled
to the Vlasov equation in $f$. Putting $t = 0$ in the Einstein
equations yields two constraint equations that link the three
initial data that are the two initial data for $\lambda(t, r)$ and 
$\mu(t, r)$ denoted $\overset{\circ}{\lambda}(r)$ and 
$\overset{\circ}{\mu}(r)$ respectively, and the initial datum for 
$f(t, r, v)$ denoted by $\overset{\circ}{f}(r, v)$. The equations to be 
solved are two first order ODE for $\overset{\circ}{\lambda}$,
$\overset{\circ}{\mu}$. The mass function $m(t,r)$ is defined as an
integral of $f$ over the hypersurfaces of constant $t$. The exact definition
is given in the next section. Provided the initial mass function
$\overset{\circ}{m}(r)= m(0,r)$ is everwhere less than $r/2$ the function
$\overset{\circ}{\lambda}$ can be determined from the relation
$1-2m/r=e^{-2\lambda}$. Once this has been done $\mu$ can be determined 
by integration. In this way functions $\overset{\circ}{\lambda}$ and 
$\overset{\circ}{\mu}$ can be determined in a straightforward way when 
$\overset{\circ}{f}$ is given, provided the inequality $2m/r<1$ is satisfied
on the initial hypersurface. Thus a simple parametrization of the set
of initial data satisfying the constraints is obtained. They can be 
constructed from a non-negative function $\overset{\circ}f$ which is
required to satisfy one inequality. 

In the case of charged particles, the problem of constraints
on the initial data is not so easy. We consider the case of a spherically 
symmetric electric field $\vec{E}$ of the form
$\vec{E}(t, r) = e(t, r)\frac{\vec{r}}{r}$, where $e(t, r)$ is an unknown 
scalar function and $\vec{r}$ the position vector in $\mathbb{R}^{3}$. 
We denote by $\overset{\circ}{e}$ the initial data
for $e(t, r)$. The Maxwell equations imply a constraint equation on the 
initial data, that is a first order ODE in the radial variable 
$r$. We now have to face the problem of constraints between the four
initial data $\overset{\circ}{\lambda}$, $\overset{\circ}{\mu}$,
$\overset{\circ}{f}$, $\overset{\circ}{e}$. If the equations for
$\overset{\circ}{\lambda}$, $\overset{\circ}{f}$, and $\overset{\circ}{e}$
can be solved then $\overset{\circ}{\mu}$ can be determined by a simple
integration as in the case of uncharged particles. On the other hand the
problem of determining the first three quantities is more difficult in 
the charged case due to the coupling of the gravitational and 
electromagnetic constraints. Any solution of the constraints must
satisfy the condition $2m/r<1$ on the initial hypersurface. However
in the charged case $m$ includes a contribution due to the energy density 
of the electric field and this depends on $e$ and $\lambda$. As a 
consequence, in contrast to the uncharged case, this condition cannot be 
expressed as an inequality on $f$ alone. Moreover, the equation for $e$
also contains $\lambda$ and so it is not possible to solve the
electromagnetic constraint separately from the gravitational one.
There is no alternative but to solve the two constraints together.
When $\overset{\circ}{f}$ is prescribed the constraints define a
system of two coupled nonlinear ordinary differential equations for
$\overset{\circ}{\lambda}$ and $\overset{\circ}{e}$. These equations
are singular at $r=0$.

Solving the ordinary differential equations arising from the constraints
is a task which must necessarily be mastered before undertaking the study
of the problem of evolution for the Einstein-Vlasov-Maxwell system.
Since the equations are singular they cannot be handled by standard results 
of ODE theory alone. In this paper, we prove, using statements developed in 
\cite{rendall2} on singular ODE, that, under smallness assumptions on the 
prescribed initial data function $\overset{\circ}{f}$, a corresponding 
global solution of the constraints exists. This provides a way of showing
the existence of initial data $\overset{\circ}{\lambda}$, 
$\overset{\circ}{\mu}$, $\overset{\circ}{e}$, in the case of an
asymptotically flat spacetime with a regular center.
In other words, we prove that the results obtained for the
constraint equations in \cite{rein} that focuses on the case of
uncharged particles, extend to the case of charged particles, in
the sense that the initial datum $\overset{\circ}{f}$ can still be
considered as the only arbitrary initial datum. 

Given the complicated nature of the constraint equations it seems
impractible to describe the most general class of functions
$\overset{\circ}{f}$ for which the constraints can be solved. 
What we have been able to do is to describe two large classes
of functions $\overset{\circ}{f}$ for which it is possible. The first
case treated is a rather obvious one to try. The charge $q$ of a 
particle is a free parameter in the equations. In the case $q=0$
the equations reduce to the uncharged situation analysed in \cite{rein}.
A perturbation argument then allows the case with fixed 
$\overset{\circ}{f}$ and small $q$ to be treated. A more exotic class
of initial data is obtained by a rigorous perturbation argument in a
regime where the density of particles tends to zero and the charge per
particle tends to infinity. This allows the construction of initial
data which coincide outside a compact set with data for the 
Reissner-Nordstr\"om solution with arbitrarily large charge to mass
ratio. The restriction of these data to the region outside a large radius 
does not allow an electrovacuum interior which has a regular centre or
corresponds to a black hole but, as this construction shows, does allow 
a regular (dynamical) interior with charged matter. 

The paper is organized as follows. In Sect. 2, we recall the spherically 
symmetric Einstein-Vlasov-Maxwell system, from which we deduce the constraint 
equations. In Sect. 3, we recall the main result of \cite{rendall2} which
we use and we prove the existence of large classes of solutions of the 
constraint equations for the Einstein-Vlasov-Maxwell system. The derivation
of some facts required in Sect. 3 concerning regions where there are no
particles is contained in an appendix.

\section{Formulation of the problem}
We consider fast moving collisionless particles with unit mass and charge
$q$. The basic spacetime is $(\mathbb{R}^{4}, g)$, with
$g$ a Lorentzian metric with signature $(-,+,+,+)$. In what
follows, we assume that Greek indices run from $0$ to $3$ and
Latin indices from $1$ to $3$, unless otherwise specified. We also
adopt the Einstein summation convention. The metric $g$ reads
locally, in cartesian coordinates $(x^{\alpha}) = (x^{0}, x^{i})
\equiv (t, \tilde{x})$:
\begin{equation} \label{eq:2.1}
ds^{2} = g_{\alpha \beta}dx^{\alpha} \otimes dx^{\beta}
\end{equation}
The assumption of spherical symmetry means that we can take $g$ of the
following form (Schwarzschild coordinates)
\begin{equation} \label{eq:2.2}
ds^{2} = - e^{2\mu}dt^{2} + e^{2\lambda}dr^{2} + r^{2}(d\theta^{2}
+ \sin^2\theta d\varphi^{2})
\end{equation}
where $\mu = \mu(t, r)$; $\lambda = \lambda(t, r)$; $t \in
\mathbb{R}$; $r \in [0, + \infty[$; $\theta \in [0,
\pi]$; $\varphi \in [0, 2\pi]$. 
The Einstein-Vlasov-Maxwell system reads:
\begin{equation} \label{eq:2.3}
R_{\alpha \beta} -\frac{1}{2}g_{\alpha \beta}R = 8\pi (T_{\alpha
\beta}(f) + \tau_{\alpha \beta}(F))
\end{equation}
\begin{equation} \label{eq:2.4}
\mathcal{L}_{X(F)} f = 0
\end{equation}
\begin{equation} \label{eq:2.5}
\nabla_{\alpha}F^{\alpha \beta} = J^{\beta}; \quad
\nabla_{\alpha}F_{\beta \gamma} + \nabla_{\beta}F_{\gamma \alpha}
+ \nabla_{\gamma}F_{\alpha \beta} = 0
\end{equation}
with:\\
\begin{align*}
T_{\alpha \beta}(f) & = - \int_{\mathbb{R}^{3}}p_{\alpha}p_{\beta}
f\omega_{p}; \quad \tau_{\alpha \beta}(F) = -
\frac{g_{\alpha\beta}}{4}F_{\gamma \nu}F^{\gamma \nu} + F_{\beta
\gamma}F_{\alpha}{}^{\gamma}\\
    J^{\beta}(f)(x) & =  q \int_{\mathbb{R}^{3}}p^{\beta}f(x, p)
\omega_{p}, \quad \omega_{p} = \mid g \mid^{\frac{1}{2}}
\frac{dp^{1}dp^{2}dp^{3}}{p_{0}}, \ p_{0} = g_{0 0}p^{0},\\
      X^{\alpha}(F) & = (p^{\alpha}, - \Gamma_{\beta
\gamma}^{\alpha}p^{\beta}p^{\gamma} - q
p^{\beta}F_{\beta}{}^{\alpha}),
\end{align*}
where $\Gamma_{\beta\gamma}^{\alpha}$ denote the Christoffel
symbols. Here, $x = (x^{\alpha})$ is the position and $p =
(p^{\alpha})$ is the 4-momentum of particles. In the expressions
above, $f$ stands for the distribution function of the charged
particles, $F$ stands for the electromagnetic field created by the
charged particles. Here (\ref{eq:2.3}) are the Einstein equations for 
the metric tensor $g = (g_{\alpha \beta})$ with sources generated by both
$f$ and $F$, that appear in the stress-energy tensor $T_{\alpha \beta} +
\tau_{\alpha \beta}$. Equation (\ref{eq:2.4}) is the Vlasov equation for 
the distribution function $f$ of the collisionless particles and 
(\ref{eq:2.5}) are the Maxwell equations for the electromagnetic
field $F$, with source (current) generated by $f$ through $J= J(f)$.
One verifies that the conservation laws $\nabla_{\alpha}(T^{\alpha
\beta} + \tau^{\alpha \beta}) = 0$ hold if $f$ satisfies the Vlasov equation.

By the assumption of spherical symmetry, we can take $g$ in the
form (\ref{eq:2.2}). One shows, using the Maxwell equation that
$F$ reduces to its electric part. We take it in the form $E =
(E^{\alpha})$ with $E^{0} = 0$, $E^{i}= e(t, r)\frac{x^{i}}{r}$,
and then, a straightforward calculation shows that: \\
\begin{align*}
\tau_{0 0} & = \frac{1}{2} e^{2(\lambda + \mu)} e^{2}(t, r);
 \quad \tau_{0i} = 0 \\
\tau_{i j} & = \frac{1}{2} e^{2\lambda} e^{2}(t, r) \left\{(\delta_{i j}
- \frac{x_{i} x_{j}}{r^{2}}) - e^{2\lambda} \frac{x_{i}
x_{j}}{r^{2}}\right\},
\end{align*}
where $\delta_{i j}$ is the Kronecker symbol.

These relations and results of \cite{rein} show that the spherically
symmetric Einstein-Vlasov-Maxwell system implies the
following first order ODE system in $\lambda$, $\mu$, $f$, $e$:
\begin{equation} \label{eq:2.6}
e^{-2\lambda}(2r\lambda' - 1) + 1 = 8\pi r^{2}\rho
\end{equation}
\begin{equation} \label{eq:2.7}
e^{-2\lambda}(2r\mu' + 1) - 1 = 8\pi r^{2}p
\end{equation}
\begin{equation} \label{eq:2.8}
\frac{\partial f}{\partial t} + e^{\mu - \lambda} \frac{v}{\sqrt{1
+ v^{2}}} \cdot\frac{\partial f}{\partial \tilde{x}} - {(e^{\mu -
\lambda} \mu' \sqrt{1 + v^{2}} + \dot{\lambda}
\frac{\tilde x\cdot v}{r} - q e^{\lambda + \mu} e(t,
r))\frac{\tilde{x}}{r}}\cdot\frac{\partial f}{\partial v} = 0
\end{equation}
\begin{equation} \label{eq:2.9}
\frac{d}{dr}(r^{2}e^{\lambda}e(t, r)) = 
q r^{2}e^{\lambda} \int_{\mathbb{R}^{3}}f(t, \tilde{x}, v)
dv
\end{equation}
where $\lambda' = \frac{\partial\lambda}{\partial r}$; \, $\dot{\lambda} =
\frac{\partial\lambda}{\partial t}$ and:
\begin{equation} \label{eq:2.10}
\rho(t, \tilde{x}) = \int_{\mathbb{R}^{3}} f(t, \tilde{x}, v)
\sqrt{1 + v^{2}} dv +
\frac{1}{2} e^{2\lambda(t, \tilde{x})} e^{2}(t, \tilde{x})
\end{equation}
\begin{equation} \label{eq:2.11}
p(t,\tilde{x}) = \int_{\mathbb{R}^{3}} (\frac{\tilde{x}\cdot v}{r})^{2}
f(t, \tilde{x}, v) \frac{dv}{\sqrt{1 + v^{2}}} - \frac{1}{2}
e^{2\lambda(t, \tilde{x})} e^{2}(t, \tilde{x})
\end{equation}
Here (\ref{eq:2.6})-(\ref{eq:2.7}) are the Einstein
equations for $\lambda$ and $\mu$, (\ref{eq:2.8}) is the
Vlasov equation for $f$ and (\ref{eq:2.9}) is the Maxwell
equation for $e$. Here $\tilde{x}$ and $v$ belong to
$\mathbb{R}^{3}$, $r := |\tilde{x}|$, $\tilde{x}\cdot v$
denotes the usual scalar product of vectors in $\mathbb{R}^{3}$, 
and $v^{2} := v\cdot v$.
The distribution function $f$ is assumed to be invariant under
simultaneous rotations of $\tilde{x}$ and $v$, hence $\rho$ and
$p$ can be regarded as functions of $t$ and $r$.
It is assumed that $f(t)$ has compact support for each fixed $t$.
We are interested in asymptotically flat spacetimes which leads to 
imposing the boundary condition that:
\begin{equation} \label{eq:2.12}
\lim_{r \to \infty}\mu(t, r)  = 0
\end{equation}
They should also have a regular centre which means that in addition
to $\lambda$, $\mu$ and $e$ being smooth functions of $t$ and $r$, including
at $r=0$, the boundary condition $\lambda(t,0)=0$ should be satisfied for
all $t$. Note that the condition that $\lambda(t,r)$ tends to zero as
$r\to\infty$, which is part of asymptotic flatness, follows from the field
equations and the fact that $f(t)$ has compact support. This is because in 
the region where $f$ vanishes the general solution of (\ref{eq:2.6}) 
satisfies the condition $\lambda(r)=O(r)$ as $r\to\infty$.

Let $m(r)=4\pi \int_0^r s^2\rho(s) ds$. This is the mass function
referred to in the introduction. Its limit $M$ as $r\to\infty$ is the
total or ADM mass of the system. The function 
$n=\int_{\mathbb{R}^{3}}f dv$ is the number density of particles and
$nq$ the charge density. The total charge of the system is given by
$Q=4\pi q\int_0^\infty s^2 e^{\lambda(s)} n(s) ds$.  

Now, define the initial data by:
\begin{equation} \label{eq:2.13}
f(0, \tilde{x}, v) = \overset{\circ}{f}(\tilde{x}, v); \,
\lambda(0, \tilde{x}) = \overset{\circ}{\lambda}(\tilde{x}) =
\overset{\circ}{\lambda}(r); \  \overset{\circ}{e}(0,\tilde{x}) =
\overset{\circ}{e}(\tilde{x}) = \overset{\circ}{e}(r)
\end{equation}
with $\overset{\circ}{f} \in C_{0}^{1}$ being a continuously
differentiable function with compact support, which is nonnegative
and spherically symmetric, i.e
\begin{displaymath}
\forall A \in SO(3), \, \forall (\tilde{x}, v) \in \mathbb{R}^{6},
\, \overset{\circ}{f}(A\tilde{x}, A v) =
\overset{\circ}{f}(\tilde{x}, v).
\end{displaymath}
We obtain the constraint equations on the initial data by taking
(\ref{eq:2.6}), (\ref{eq:2.7}) and (\ref{eq:2.9}) for $t=0$, that
give:
\begin{equation} \label{eq:2.14}
e^{-2\overset{\circ}{\lambda}}(2r\overset{\circ}{\lambda}' - 1) +
1 = 8\pi r^{2} \int_{\mathbb{R}^{3}} \sqrt{1 + v^{2}}
\overset{\circ}{f}(r, v) dv + 4\pi r^{2}\,
e^{2\overset{\circ}{\lambda}}\, \overset{\circ}{e}^{2}
\end{equation}
\begin{equation} \label{eq:2.15}
\frac{d}{dr}(r^{2}e^{\overset{\circ}{\lambda}}\overset{\circ}{e})
= q r^{2}e^{\overset{\circ}{\lambda}}\int_{\mathbb{R}^{3}}
\overset{\circ}{f}(r, v) dv = J(\overset{\circ}{f})
\end{equation}
\begin{equation} \label{eq:2.16}
e^{-2\overset{\circ}{\lambda}}(2r\overset{\circ}{\mu}' + 1) - 1 =
8\pi r^{2} \int_{\mathbb{R}^{3}} (\frac{\tilde x\cdot v}{r})^{2}
\overset{\circ}{f}(r, v) \frac{dv}{\sqrt{1 + v^{2}}} -
4\pi r^{2}\, e^{2\overset{\circ}{\lambda}}\,
\overset{\circ}{e}^{2}
\end{equation}
We observe that, if $\overset{\circ}{f}$ is given and if we can
solve (\ref{eq:2.14})-(\ref{eq:2.15}) for
$\overset{\circ}{\lambda}$ and $\overset{\circ}{e}$ then
(\ref{eq:2.16}) determines at once $\overset{\circ}{\mu}'$. Using the
boundary condition (\ref{eq:2.12}) then determines $\overset{\circ}{\mu}$.
So we can concentrate on (\ref{eq:2.14})-(\ref{eq:2.15}).
In what follows, we fix $\overset{\circ}{f}$ in
(\ref{eq:2.14})-(\ref{eq:2.15}) and we look for a unique global
asymptotically flat solution $(\overset{\circ}{\lambda}, \overset{\circ}{e})$
of the system (\ref{eq:2.14})-(\ref{eq:2.15}) above with regular centre.
Note that, using the compact support of $\overset{\circ}{f}$ and the 
equations (\ref{eq:2.14}) and (\ref{eq:2.15}), it follows that 
$\overset{\circ}{\lambda}$ and $\overset{\circ}{e}$ tend to zero as 
$r\to\infty$. It also follows from (\ref{eq:2.14}) and (\ref{eq:2.15}) and 
the regularity of the solution that $\overset{\circ}{\lambda}'(0)=0$ and 
$\overset{\circ}{e}(0)=0$.

\section{Existence of global solutions of the constraints}

In this section the existence of global solutions of the equations
(\ref{eq:2.14}) and (\ref{eq:2.15}) will be proved. Let us state first 
of all the following result of \cite{rendall2} on which our global 
existence theorem relies:
\begin{theorem} \label{T:3.1}
Let $V$ be a finite-dimensional real vector space, $N : V
\rightarrow V$ a linear mapping, $G : I \times V \rightarrow V$ a
smooth (i.e $C^{\infty}$) mapping and $g : I \rightarrow V$ a
smooth mapping, where $I$ is an open interval in $\mathbb{R}$
containing zero. Consider the equation
\begin{equation} \label{eq:3.1}
s\frac{df}{ds} + Nf = sG(s, f(s)) + g(s)
\end{equation}
for a function $f$ defined on a neighborhood of $0$ in $I$ and
taking values in $V$. Suppose that each eigenvalue of $N$ has a
positive real part. Then there exists an open interval $J$ with $0
\in J \subset I$ and a unique bounded $C^{1}$ function $f$ on $J
\setminus \{ 0 \}$ satisfying (\ref{eq:3.1}). Moreover $f$ extends
to a $C^{\infty}$ solution of (\ref{eq:3.1}) on $J$. If $N$, $G$
and $g$ depend smoothly on a parameter $z$ and if the eigenvalues
of $N$ are distinct then the solution also depends smoothly on
$z$.
\end{theorem}
\textbf{Proof}. See Theorem 1 in \cite{rendall2}, p.989.
\begin{remark}
The assumption that $N$ has distinct eigenvalues is to ensure that
$N$ can be reduced to diagonal form by a similarity transformation
depending smoothly on $z$. In particular, Theorem 3.1
applies if $N$ is already a diagonal matrix.
\end{remark}
\begin{theorem}[Local existence] \label{T:3.2}
Let $\overset{\circ}{f} \in C^\infty(\mathbb{R}^{6})$ be
nonnegative, compactly supported and spherically symmetric. Then, the 
equations (\ref{eq:2.14}) and (\ref{eq:2.15}) have a
unique local and regular solution $(\overset{\circ}{\lambda},
\overset{\circ}{e})$ defined on some interval $[0, R]$, $R > 0$. The
solution depends smoothly on the parameter $q$.
\end{theorem}
\textbf{Proof}: Let $\overset{\circ}{f} \in
C^\infty(\mathbb{R}^{6})$ be nonnegative, compactly supported and 
spherically symmetric. By a regular solution we mean one which is smooth
and for which $\overset{\circ}{\lambda}=0$ at $r=0$. It follows that for
any regular solution  $\overset{\circ}{\lambda}$ can be written in the
form:
\begin{equation} \label{eq:3.2}
\overset{\circ}{\lambda}(r) = rL(r)
\end{equation}
for a smooth function $L(r)$. Equation (\ref{eq:3.2}) implies 
$\overset{\circ}{\lambda}' = L + r L'$ and
(\ref{eq:2.14}) - (\ref{eq:2.15}) can be written:
\begin{equation} \label{eq:3.3}
rL' + L = \frac{1}{2r}(1 - e^{2rL}) +
4\pi r e^{2rL}(\int_{\mathbb{R}^{3}} \sqrt{1+v^{2}}
\overset{\circ}{f}(r, v) dv +
\frac{1}{2}e^{2rL}\overset{\circ}{e}^{2})
\end{equation}
\begin{equation} \label{eq:3.4}
r \overset{\circ}{e}' + 2 \overset{\circ}{e} = -r
\overset{\circ}{e}(L + rL') + r q \int_{\mathbb{R}^{3}}
\overset{\circ}{f}(r, v) dv
\end{equation}
The function $e^{2x}-1-2x$ vanishes to first order at the origin and 
hence $e^{2x}-1-2x=x^2F_0(x)$ for a smooth function $F_0$. Hence 
the equation for $L$ can be rewritten in the form
\begin{equation*}
rL' + L = -L + \frac{r}{2}L^{2}F_{0}(rL) +
4\pi r e^{2rL}(\int_{\mathbb{R}^{3}} \sqrt{1+v^{2}}
\overset{\circ}{f} dv +
\frac{1}{2}e^{2rL} \overset{\circ}{e}^{2})
\end{equation*}
Thus
\begin{equation} \label{eq:3.5}
rL' + 2L = rG_{1}(r, L, \overset{\circ}{e}, \overset{\circ}{f})
\end{equation}
where
\begin{equation*}
G_{1}(r, L, \overset{\circ}{e}, \overset{\circ}{f}) =
\frac{1}{2}L^{2}F_{0}(rL) + 4\pi e^{2rL}
\int_{\mathbb{R}^{3}}\sqrt{1 +v^{2}} \overset{\circ}{f}(r, v) dv +
2\pi e^{4rL} \overset{\circ}{e}^{2}
\end{equation*}
and (\ref{eq:3.4}) reads, given (\ref{eq:3.5}):
\begin{equation*}
r\overset{\circ}{e}' + 2\overset{\circ}{e} =
-r\overset{\circ}{e}(-L + rG_{1}(r, L, \overset{\circ}{e},
\overset{\circ}{f})) + r q
\int_{\mathbb{R}^{3}}\overset{\circ}{f}(r, v) dv.
\end{equation*}
Hence:
\begin{equation} \label{eq:3.6}
r\overset{\circ}{e}' + 2\overset{\circ}{e} = rG_{2}(r, L,
\overset{\circ}{e}, \overset{\circ}{f})
\end{equation}
where
\begin{equation*}
G_{2}(r, L, \overset{\circ}{e}, \overset{\circ}{f})=
L\overset{\circ}{e} - r\overset{\circ}{e}G_{1}(r, L,
\overset{\circ}{e}, \overset{\circ}{f}) +q
\int_{\mathbb{R}^{3}}\overset{\circ}{f}(r, v) dv
\end{equation*}
Setting $G =
\begin{pmatrix}
G_{1} \\
G_{2}
\end{pmatrix}$ and $\Phi =
\begin{pmatrix}
L \\
\overset{\circ}{e}
\end{pmatrix}$ and using (\ref{eq:3.5})-(\ref{eq:3.6}), the equations
(\ref{eq:2.14}) and (\ref{eq:2.15})
can be written:
\begin{equation} \label{eq:3.7}
r\frac{d\Phi}{dr} + 2\Phi = rG(r,\Phi(r))
\end{equation}
We apply Theorem 3.1 with $V = \mathbb{R}^{2}$, $N =
\begin{pmatrix}
 2 & 0\\
 0 & 2
\end{pmatrix}$, $N \Phi = 2 \Phi$ to (\ref{eq:3.7}) and, since $G$ clearly 
depends smoothly on $q$, obtain the desired result. Thus Theorem 3.2 is 
proved.

\begin{theorem}[Global existence, low charge] \label{T:3.3}
Let $\overset{\circ}{f}\in C^\infty(\mathbb{R}^{6})$ be
nonnegative, compactly supported and spherically symmetric with
\begin{equation} \label{eq:3.8}
8\pi \int_{0}^{r} s^{2}(\int_{\mathbb{R}^{3}}\sqrt{1+v^2}
\overset{\circ}{f}(s,v)
dv)ds < r.
\end{equation}
Then, for $q$ small enough, the equations
(\ref{eq:2.14}) and (\ref{eq:2.15}) have a unique
global and regular solution $(\overset{\circ}{\lambda},
\overset{\circ}{e})$ defined on $[0, + \infty[$ that satisfies the
boundary condition $\overset{\circ}{\lambda}(0)=0$.
\end{theorem}
\textbf{Proof} Let $\overset{\circ}{f} \in
C^\infty(\mathbb{R}^{6})$ be nonnegative, compactly supported and 
spherically symmetric. We assume that $\overset{\circ}{f}$ is fixed and 
satisfies (\ref{eq:3.8}). By Theorem
3.2, the equations (\ref{eq:2.14}) and (\ref{eq:2.15}) have a unique
local regular solution on some interval $[0, R]$, $R > 0$.
Again, Theorem 3.1 shows that, for fixed $\overset{\circ}{f}$,
there exists $E > 0$, such that for $q \in [-E, E]$, $R$
can be chosen uniformly and the solution on $[0, R]$ depends
continuously on the parameter $q$. Now, for fixed
$\overset{\circ}{f}$ and $q$, the solution has a
right maximal interval of existence $[0, R_{\ast}[$, $R_{\ast} =
R_{\ast}(\overset{\circ}{f}, q)$. We have to show
that $R_{\ast} = + \infty$.
In fact, the second term in the r.h.s of (\ref{eq:2.14}) vanishes
for $q = 0$, as one can see by integrating (\ref{eq:2.15})
over $[0, r]$, $r > 0$. It follows that for $q = 0$,
(\ref{eq:2.14}) and (\ref{eq:2.15}) have a global solution under the sole
assumption (\ref{eq:3.8}) on $\overset{\circ}{f}$.
Then by the stability theorem for ODE, for every $R > 0$, there
exists a number $E > 0$, such that, for every $q \in [-E,
E]$, the system (\ref{eq:3.7}) has a solution $\Phi_{E}$ that
exists on $[0, R]$ (see Theorem 4, p. 92 in \cite{perko}). Thus
$R_{\ast} > R$. Now, we can choose $R$ large so that ${\rm supp}
\overset{\circ}{f} \subset [0, R] \times {\mathbb{R}^{3}}$, i.e
$\overset{\circ}{f}(r, v) = 0$ for $r \geq R$. If $R_0$ is the radius
of the support of the distribution function then $R$ may be chosen to
be bigger than $m(R_0)+Q^2/(8\pi R_0)$ for all $q$ in the interval
$[E,E]$. Hence by the lemma of the appendix the solution extends to
one which is global and regular. This completes the proof of the 
theorem.

\begin{theorem}[Global existence, high charge] 
Let $\bar f\in C^\infty(\mathbb{R}^{6})$ be
nonnegative, compactly supported and spherically symmetric. 
Then, for $q$ large enough, the equations (\ref{eq:2.14}) and (\ref{eq:2.15}) 
have a global and regular solution $(\overset{\circ}{\lambda},
\overset{\circ}{e})$ defined on $[0, + \infty[$ that satisfies the
boundary condition $\overset{\circ}{\lambda}(0)=0$ for which 
$\overset{\circ}{f}$ is a constant multiple of $\bar f$. Moreover the charge 
to mass ratio $Q/M$ of the solution can be made as large as desired.
\end{theorem}
\textbf{Proof}
We assume that $\bar f$ is fixed as in the assumptions of the theorem
We set:
\begin{equation*}
\alpha = q^{-1}; \quad \bar{f} =
\alpha^{-k}\overset{\circ}{f}; \quad \bar{e} = \alpha^{-(k -
1)}\overset{\circ}{e}
\end{equation*}
for some integer $k \geq 2$. Then (\ref{eq:2.14})-(\ref{eq:2.15})
can be written as:
\begin{equation} \label{eq:3.9}
e^{-2\overset{\circ}{\lambda}}(2r\overset{\circ}{\lambda}' - 1) +
1 = 8\pi r^{2}(\alpha^{k} \int_{\mathbb{R}^{3}} \bar{f}(r, v) dv +
\frac{1}{2}e^{2\overset{\circ}{\lambda}}\alpha^{2(k -
1)}\bar{e}^{2})
\end{equation}
\begin{equation} \label{eq:3.10}
2\bar{e} + r\bar{e}\overset{\circ}{\lambda}' + r\bar{e}' =
-r\int_{\mathbb{R}^{3}} \bar{f}(r, v) dv.
\end{equation}
Introducing a variable $L$ as defined in (\ref{eq:3.2}) puts these
equations into a form closely analogous to that obtained in the
proof of the last theorem. In fact the left hand side has the same 
form as in that case. All that is changed is the form of the nonlinear 
terms on the right hand side. The equations depend on $\alpha$ as a 
parameter in a way which is smooth in a neighbourhood of  $\alpha=0$. 
For $\alpha=0$ the function $\overset{\circ}{\lambda}$ vanishes identically 
while the equation for $\bar e$ becomes linear and has a global
regular solution. From this point on we can argue just as in the proof 
of Theorem 3.3 to conclude that for $\alpha$ sufficiently small there is 
a unique global regular solution of these equations. Here we must use the 
fact that $m(R_0)+Q^2/(8\pi R_0)$ is bounded independently of $\alpha$ for
$\alpha$ small. The assumption that $\alpha$ is small corresponds
to $q$ being large. The distribution function belonging to the
solution is obtained from $\overset{\circ}{f}$ by a constant rescaling.
The charge to mass ratio of the solution is proportional to 
$\alpha^{-k}$ and thus tends to infinity as $\alpha$ tends to zero.
This completes the proof.

\begin{remark}
The solution in the exterior region is part of the Reissner-Nordstr\"om
solution \cite{hawking}. 
\end{remark}

\begin{remark}
Our motivation in proving these theorems was to construct initial data
for the Einstein-Vlasov-Maxwell system. The same arguments apply with
other kinds of charged matter such as a charged fluid as sources for
the Einstein equations.
\end{remark}

\vskip 10pt\noindent
\textbf{Acknowledgments}: One of us (ADR) thanks Curt Cutler for a
helpful suggestion. This work was supported by a research grant from
the VolkswagenStiftung, Federal Republic of Germany. 

\appendix
\section{Appendix}
This appendix contains an analysis of exterior regions free of particles in 
initial data sets for the Einstein-Vlasov-Maxwell system. Let 
$\overset{\circ}{f}$ be an initial distribution function of compact
support. Let $R_0$ be the radius of its support in space so that 
$\overset{\circ}{f}$ vanishes for all $r\ge R_0$. In this appendix 
we are only concerned with quantities on the initial hypersurface and so
we will drop the label zero indicating the restriction of spacetime
quantities to the initial hypersurface. 

\begin{lemma}
Consider a solution of the constraint equations for the spherically
symmetric Einstein-Vlasov-Maxwell system defined for $0\le r\le R_1$
and having a regular centre. Suppose that radius $R_0$ of the support of
the distribution function $f$ is less than $R_1$. Let 
$\tilde M=m(R_0)+Q^2/(8\pi R_0)$. Then if $R_1>2\tilde M$ the given 
solution extends to a unique solution of the constraints defined for all 
$r\in [0,\infty[$ which is asymptotically flat and has $f=0$ for 
$R\ge R_0$.
\end{lemma}

\noindent
\textbf{Proof}
Integrating the constraint equation (\ref{eq:2.15}) gives
\begin{equation}
r^2e^{\lambda(r)}e(r)=q\int_0^r s^2e^{\lambda(s)}\int_{\mathbb{R}^{3}}
f(s,v) dv
\end{equation}
For $r\ge R_0$ the upper limit $r$ in the integral can be replaced by 
$R_0$ or infinity without changing the value of the expression. It
is equal to $Q/4\pi$ where $Q$ is the total charge of the system defined
in Sect. 2. For $r\ge R_0$ the function $f$  
vanishes and the mass function $m$ defined in Sect. 2 satisfies
\begin{equation}
m'=\frac{2\pi}{r^2}(Q/4\pi)^2
\end{equation}
It follows that $\tilde M(r)=m(r)+Q^2/(8\pi r)$ is independent of $r$.
If the solution exists globally in $r$ and is asymptotically flat then
taking the limit $r\to\infty$ shows that $\tilde M$  is equal to the ADM 
mass $M$. In any case $\tilde M$ is positive and it follows that
in the exterior region $m=\tilde M-Q^2/(8\pi r)$. In order to 
determine whether a solution can be extended to larger values of
the radius it is enough to ensure that $1-2\tilde M/r+Q^2/(4\pi r^2)$ remains
positive. For in that case we can define $\lambda$ by means of the relation
\begin{equation}
e^{-2\lambda}=1-2\tilde M/r+Q^2/(4\pi r^2)
\end{equation}
Note that $\lim_{r\to\infty}\lambda(r)=0$.
Once this has been done we can define $\mu$ to be equal to $-\lambda$
and $e(r)=r^{-2}e^{-\lambda}(Q/4\pi)$ in the external region and this 
gives the unique solution satisfying the correct boundary conditions.
If $r>2\tilde M$ then $1-2\tilde M/r+Q^2/(8\pi r^2)$ is automatically
positive and the desired result is obtained.

\end{document}